\documentclass[pre,twocolumn,superscriptaddress,preprintnumbers,aps,showpacs,amsmath,amssymb]{revtex4}

\usepackage{graphicx}
\usepackage{dcolumn}
\usepackage{bm}
\usepackage{color}

\begin{document}

\title{Oscillation of a Rotating Levitated Droplet: Analysis with a Mechanical Model}

\author{Hiroyuki Kitahata\footnote{Corresponding author. E-mail: kitahata@chiba-u.jp.}}
\affiliation{Department of Physics, Graduate School of Science, Chiba University, Chiba 263-8522, Japan}

\author{Rui Tanaka}
\affiliation{Graduate School of System and Information Engineering, University of Tsukuba, Tsukuba, Ibaraki, 305-8573, Japan}

\author{Yuki Koyano}
\affiliation{Department of Physics, Graduate School of Science, Chiba University, Chiba 263-8522, Japan}

\author{Satoshi Matsumoto}
\affiliation{Institute of Space and Astronautical Science, Japan Aerospace Exploration Agency, Tsukuba, Ibaraki 305-8505, Japan}

\author{Katsuhiro Nishinari}
\affiliation{Research Center for Advanced Science and Technology, The University of Tokyo,
Meguro-ku, Tokyo 153-8904, Japan}

\author{Tadashi Watanabe}
\affiliation{Research Institute of Nuclear Engineering, University of Fukui, Tsuruga, Fukui 914-0055, Japan}

\author{Koji Hasegawa}
\affiliation{Faculty of Engineering, Kogakuin University, Shinjuku-ku, Tokyo 163-8677, Japan}

\author{Tetsuya Kanagawa}
\affiliation{Graduate School of System and Information Engineering, University of Tsukuba, Tsukuba, Ibaraki, 305-8573, Japan}

\author{Akiko Kaneko}
\affiliation{Graduate School of System and Information Engineering, University of Tsukuba, Tsukuba, Ibaraki, 305-8573, Japan}

\author{Yutaka Abe}
\affiliation{Graduate School of System and Information Engineering, University of Tsukuba, Tsukuba, Ibaraki, 305-8573, Japan}

\begin{abstract}
A droplet of millimeter-to-centimeter scale can exhibit electrostatic levitation, and such levitated droplets can be used for the measurement of the surface tension of the liquids by observing the characteristic frequency of oscillatory deformation. In the present study, a simple mechanical model is proposed by considering a single mode of oscillation in the ellipsoidal deformation of a levitated rotating droplet. By measuring the oscillation frequency with respect to the rotational speed and oscillation amplitude, it is expected that the accuracy of the surface tension measurement could be improved. Using the proposed model, the dependences of the characteristic frequency of oscillatory deformation and the averaged aspect ratio are calculated with respect to the rotational angular velocity of a rotating droplet. These dependences are found to be consistent with the experimental observations.
\end{abstract}

\pacs{82.40.Bj, 05.45.-a, 47.55.D-}

\maketitle

\section{Introduction}

The dynamics of a droplet is one of the most interesting topics in physics~\cite{DynamicsDroplet}. In the 19th century, Rayleigh investigated the frequency of the fundamental mode of oscillation in droplet shape~\cite{Rayleigh}. Since then, there have been several studies on the dynamics of droplets, for example, on the dynamics of the collision of a droplet~\cite{RibouxPRL,KlaseboerPRL} and the wetting of a droplet on a substrate~\cite{LeePRE,OkumuraEPL}. There have also been many studies on the oscillation of droplet shape under various conditions~\cite{OhsakaPRL,DuftPRL,CourtyPRE}.

Through such studies, the techniques for levitating a droplet have been improved. A droplet of the millimeter-to-centimeter scale can be levitated in a static electric field by adding an electric charge to it; this phenomenon is called electrostatic levitation. By using such a levitated droplet, new materials can be synthesized without disturbance by a container wall. Levitated droplets have also been utilized for the measurement of physical properties of liquids such as surface tension and viscosity, without using a container~\cite{DynamicsDroplet,Ishikawa}. 

The surface tension of a liquid can be determined by measuring the characteristic frequency of an infinitesimally small oscillation in the shape of a levitating droplet of the liquid, induced by a small external driving force~\cite{Rayleigh,Landau}. The characteristic frequency of oscillation in the $n$-mode deformation was also calculated as a natural extension of the ellipsoidal deformation~\cite{Lamb}. Moreover, Busse analyzed the characteristic frequency of a rotating droplet and showed that the frequency is a decreasing function of the rotational speed as long as the amplitude of the deformation is small~\cite{Busse}. This tendency was confirmed experimentally by Annamalai {\it et al.}~\cite{Annamalai}.

On the other hand, when the amplitude of the oscillation in droplet deformation is not small, it was experimentally observed that the characteristic frequency increased as the amplitude increased in the case without rotation~\cite{Trinh-Wang,Becker}. The dependence of the amplitude was analytically estimated, and the analytical predictions were found to be consistent with the experimental results~\cite{Tsamopoulos}. Recently, some of the authors of this paper reported their experimental observations on the dependence of the characteristic frequency of oscillatory deformation for a rotating droplet with large amplitude~\cite{Tanaka}. Their results were reproduced numerically through hydrodynamic analyses~\cite{Watanabe,Watanabe2}. However, there has been no analytical studies on the characteristic frequency of a rotating droplet with large deformation.

\begin{figure}
\begin{center}
\includegraphics{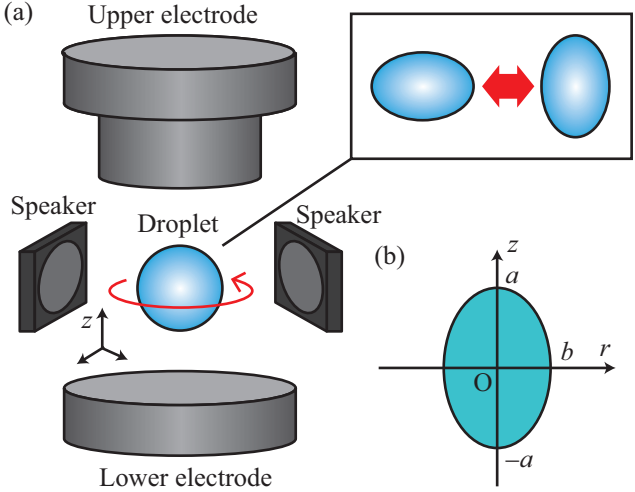}
\end{center}
\caption{(Color online)
(a) Schematic illustration of experimental setup. The droplet levitated by the static electric field is rotated by the acoustic field, and the sinusoidal modulation of the electric field induces the oscillation in the ellipsoidal deformation. (b) Definition of variables, $a$ and $b$, representing deformation.}
\label{fig1}
\end{figure}

In the present work, the dependence of the characteristic frequency of oscillatory deformation is investigated for a levitated rotating droplet with large amplitude, based on a simple mechanical model. The study focuses on a single fundamental mode of deformation, {\it i.e.}, ellipsoidal deformation, and analyzes the properties of the oscillation of the levitated rotating droplet. Then the analytical results are compared with experimental observations. The corresponding experimental setup is illustrated in Fig.~\ref{fig1}(a). By adding a positive charge to the droplet and applying a static electric field between the upper and lower electrodes, the droplet is levitated against gravity. A sinusoidal modulation of the voltage between the two electrodes then induces an oscillation in the ellipsoidal deformation of the droplet. Finally, a torque is exerted on the droplet by applying an acoustic field using two speakers located orthogonally to each other, and the droplet rotates. By measuring the dependence of the frequency on the amplitude and rotational speed, it is expected that the accuracy of the surface tension measurement will be improved.

\section{Model}

The mechanical model is constructed under the following three assumptions: (i) A levitated droplet has an ellipsoidal shape with a constant volume. (ii) The fluid inside the droplet moves elastically; in other words, the fluid particles do not change their configuration. (iii) There is no energy dissipation.

Here, the cylindrical coordinates are adopted such that the symmetric axis corresponds to the $z$-axis in Fig.~\ref{fig1}. The shape of the ellipsoid is then described as 
\begin{equation}
\frac{r^2}{b^2}+\frac{z^2}{a^2} = 1,
\end{equation}
which can also be written as
\begin{equation}
r(z) = \frac{b}{a} \sqrt{a^2 - z^2},
\end{equation}
where $a$ and $b$ are positive, as illustrated in Fig.~\ref{fig1}(b).

The volume $V$ and surface area $S$ of the ellipsoid are given by
\begin{equation}
V =  \frac{4 \pi}{3} a b^2, 
\end{equation}
and
\begin{equation}
S  
= \left\{ \begin{array}{ll} \displaystyle{ 2 b^2 \pi \left( 1 + \frac{ {\rm arcsin} \sqrt{1 - \kappa^2}}{\kappa \sqrt{1 - \kappa^2}} \right)}, & {\rm for} \;\kappa \leq 1, \\ \displaystyle{2 b^2 \pi \left( 1 + \frac{{\rm arcsinh} \sqrt{\kappa^2 - 1}}{\kappa \sqrt{\kappa^2 - 1}} \right), }& {\rm for} \; \kappa \geq 1, \end{array} \right.
\end{equation}
where $\kappa = b/a$.
Because the volume is conserved, $b$ is a function of $a$, {\it i.e.}, $b = \sqrt{R^3/a}$, where $R$ is the equivalent-volume radius satisfying $4 \pi R^3 / 3 = V$. As a result, the shape of the ellipsoidal droplet is represented by only one parameter, $a$. In order to consider the deformation from a spherical shape, a new nondimensionalized variable $\xi$ is defined as
\begin{equation}
a = R \left(1 + \xi \right),
\end{equation}
where $\xi = 0$ represents a perfectly spherical shape. In addition, positive and negative $\xi$ correspond to a prolate and an oblate, respectively.
The surface area of the droplet can be expanded with respect to $\xi$ as
\begin{align}
S =& 4 \pi R^2 \left( 1 + \frac{2}{5} \xi^2 - \frac{52}{105}\xi^3 \right. \nonumber \\ &\left. + \frac{11}{21} \xi^4 -\frac{608}{1155} \xi^5 \right) + \mathcal{O}\left( \xi^6 \right), \label{surface}
\end{align}
where $\mathcal{O}(\xi^n)$ denotes the terms of the same order as or higher order than $\xi^n$.

In order to derive the equation of motion, the Lagrangian, $\mathcal{L}$ is considered as a function of $a$ and $\theta$, where $\theta$ is the characteristic direction of the droplet, introduced to describe the droplet rotation.
The Lagrangian $\mathcal{L}$ can be written as
\begin{equation}
\mathcal{L} = E_{\rm k} - E_{\rm s},
\end{equation}
where $E_{\rm k}$ and $E_{\rm s}$ are the kinetic energy and potential energy, respectively.
Under the above-mentioned assumption, the velocity of a fluid particle located at $(r, \theta, z)$ is written as $( \dot{b} r/b , r \dot{\theta} , \dot{a} z/a )$, where $\dot{a}$, $\dot{b}$, and $\dot{\theta}$ are the time derivatives of $a$, $b$, and $\theta$, respectively. It can be confirmed that the flow profile satisfies the Navier-Stokes equation (see Appendix~\ref{ap-NS}).  Thus, the kinetic energy is calculated as 
\begin{equation}
E_{\rm k} = \frac{\pi}{15} \rho \left(2 R^3 + \frac{R^6}{a^3} \right){\dot{a}}^2 + \frac{4}{15} \pi \rho \dot{\theta}^2 \frac{R^6}{a},
\end{equation}
where $\rho$ is the density of the fluid inside the droplet.
To evaluate $E_{\rm s}$, only the surface energy of the droplet originating from the surface tension is considered:
\begin{equation}
E_{\rm s} = \gamma S(a),
\end{equation}
where $\gamma$ is the surface tension of the fluid. Therefore, the Lagrangian can be written as
\begin{equation}
\mathcal{L} = \frac{\pi}{15} \rho \left(2 R^3 + \frac{R^6}{a^3} \right){\dot{a}}^2 + \frac{4}{15} \pi \rho \dot{\theta}^2 \frac{R^6}{a} - \gamma S(a).
\end{equation}

The Euler-Lagrange equation for $\theta$ represents the angular momentum conservation, {\it i.e.}, $\dot{\theta}/a = \Omega / R$, where $\Omega$ is a parameter corresponding to the initial angular momentum. By substituting this relation in the Euler-Lagrange equation for $a$, a second-order ordinary differential equation (ODE) for $a$ is derived as
\begin{equation}
\frac{2\pi \rho}{15} \left(2 R^3 + \frac{R^6}{a^3} \right) \ddot{a} - \frac{\pi \rho}{5} \frac{R^6}{a^4} \dot{a}^2  + \frac{4\pi \rho}{15} \Omega^2 R^4 + \gamma \frac{\partial S}{\partial a} = 0. \label{equ}
\end{equation}
This is the governing equation of the dynamics of the droplet.
In order to consider the deformation from a spherical shape, eq.~(\ref{equ}) can be rewritten as an ODE for $\xi$:
\begin{align}
&\ddot{\xi} + 8 \Gamma \xi \left(1 - \frac{6}{7}\xi -\frac{5}{21} \xi^2 + \frac{117}{77} \xi^3 \right) - \frac{1 - 3\xi + 5\xi^2}{2} \dot{\xi}^2  \nonumber \\ &+ \frac{2}{3} \Omega^2 \left(1 + \xi - \xi^2+ \frac{1}{3} \xi^3 +\frac{2}{3} \xi^4 \right)  = 0, \label{xi}
\end{align}
where $\Gamma = \gamma / (R^3 \rho)$, and the terms higher than the fourth-order terms of $\xi$ and $\dot{\xi}$, and the second-order terms of $\ddot{\xi}$ are neglected. Hereafter, eq.~(\ref{xi}) is analyzed for understanding the dynamics of deformation of a rotating droplet.

\section{Analysis}

Linearizing eq.~(\ref{xi}) with respect to $\xi$,
\begin{equation}
\ddot{\xi} + 8 \Gamma \xi + \frac{2}{3}\Omega^2 (1 + \xi) = 0.
\end{equation}
Thus, the characteristic frequency, $f_0$, of the oscillation in droplet deformation without rotation, {\it i.e.}, for $\Omega = 0$, is calculated as
\begin{equation}
f_0 = \frac{1}{2 \pi} \sqrt{8 \Gamma} = \frac{1}{\pi}\sqrt{\frac{2 \gamma}{\rho R^3}},
\end{equation}
which corresponds to the result obtained by Rayleigh~\cite{Rayleigh}. When the droplet rotates, $\xi = 0$ is no longer at equilibrium, but the equilibrium state shifts to $\xi = \xi_0$, which is explicitly written as
\begin{equation}
\xi_0 = -\frac{\Omega^2}{12\Gamma} + \mathcal{O}\left(\Omega^4\right),
\end{equation}
where $\Omega$ is regarded as a small parameter. Then, the characteristic frequency of the droplet oscillation, $f$, is formulated as
\begin{align}
f &= f_0 \left(1 + \frac{19 \Omega^2}{168 \Gamma} \right)+ \mathcal{O}\left(\Omega^4\right) \nonumber \\ &= f_0 \left(1 + \frac{19}{21} \frac{\Omega^2}{{\omega_0}^2}\right)+ \mathcal{O}\left(\Omega^4\right),
\end{align}
where $\omega_0$ is the angular velocity of the droplet oscillation without rotation, {\it i.e.}, $\omega_0 = 2 \pi f_0 = \sqrt{8 \Gamma}$. These results are comparable with those reported in the studies by Busse~\cite{Busse} and Annamalai {\it et al.}~\cite{Annamalai}.

To consider the nonlinear effect of large deformation on the frequency of oscillatory deformation of the droplet, a weakly nonlinear analysis is conducted~\cite{Strogatz}. As eq.~(\ref{xi}) has a form similar to eq.~(\ref{nonlin}) in Appendix~\ref{ap-derivation}, the frequency and amplitude can be calculated by eqs.~(\ref{eq-freq})~and~(\ref{eq-shift}), respectively. 
Comparing eq.~(\ref{xi}) with eq.~(\ref{nonlin}), $\alpha_1 = -1/2 + 3 \xi_0/2$, $\alpha_2 = 3/2 -5\xi_0$, $\alpha_3/{\omega_0}^2 = -6/7 - 100 \xi_0/49$, and $\alpha_4/{\omega_0}^2 = -5/21 + 2748 \xi_0/539$. Thus, the frequency, $f$, with respect to the angular velocity of the rotation, $\Omega$, and the amplitude of $\xi$ for the oscillation, $A$, can be calculated as
\begin{equation}
f= f_0 \left(1 + \frac{19}{21} \frac{\Omega^2}{{\omega_0}^2} - \frac{379}{784} A^2 \right) + \mathcal{O}\left(\Omega^4, \Omega^2A^2, A^4\right). \label{freq}
\end{equation}
The averaged aspect ratio, $\left<{\rm Ar} \right>$ is defined as
\begin{equation}
\left<{\rm Ar} \right> = \frac{\left< b \right>}{\left< a \right>},
\end{equation} 
where $\left< a \right>$ and $\left< b \right>$ are the center positions of the oscillation in the long and short axes, $a$ and $b$, respectively. Therefore, the time series of $a$ and $b$ are assumed to be $a = \left< a \right> + a_1 \cos (2 \pi f t + \delta)$ and $b = \left< b \right> - b_1 \cos (2 \pi f t + \delta)$, where $a_1$ and $b_1$ are the amplitudes and $\delta$ is the phase shift. It should be noted that $a_1$ corresponds to $R A$. The center positions can be written explicitly as
\begin{equation}
\left< a \right> = R \left(1 - \frac{2}{3} \frac{\Omega^2}{{\omega_0}^2} + \frac{19}{28} A^2 \right) + \mathcal{O}\left(\Omega^4, \Omega^2 A^2, A^4 \right),
\end{equation}
and 
\begin{align}
\left< b \right> &= \sqrt{R^3 / \left< a \right>} \nonumber \\
&= R \left(1 + \frac{1}{3}  \frac{\Omega^2}{{\omega_0}^2} - \frac{19}{56} A^2 \right)  + \mathcal{O}\left(\Omega^4, \Omega^2A^2, A^4 \right).
\end{align}
Therefore, $\left<{\rm Ar} \right>$ is obtained as
\begin{equation}
\left< {\rm Ar} \right> = \sqrt{\frac{R^3}{\left< a \right>^3}} = 1 + \frac{\Omega^2}{{\omega_0}^2}  - \frac{57}{56}A^2  + \mathcal{O}\left(\Omega^4, \Omega^2A^2, A^4 \right). \label{ar}
\end{equation}

Tsamopoulos and Brown reported the frequency shift due to the large amplitude in the ellipsoidal deformation as $f = f_0 (1 - (34409/ 58800) A^2)$~\cite{Tsamopoulos,Tsamopoulos2,Wang}.
This result is different from the one obtained in the above discussion (eq.~(\ref{freq})) because Tsamopoulos and Brown investigated the large two-mode deformation from a perfect sphere, whereas we consider an ellipsoidal droplet is considered in this study. These two approaches provide the same results for infinitesimally small deformations, but they exhibit deviations when large deformations are considered.

Next, the results obtained above are compared with a simple spring-bead model. The simplest model that can describe the deformation and rotation is a system with two equivalent small beads connected by a spring on a two-dimensional plane. The mass of each bead is set as $m$, and the potential for the extension of the spring is set as $U(\ell)$, where $\ell$ is the distance between the two beads. It is noted that $U(\ell) = k(\ell - 2\ell_0)^2/2$, when the spring responds linearly. Here, the natural length of the spring is $2\ell_0$, and the spring constant is $k$. In this case the frequency, $f'$, is obtained as
\begin{equation}
f' = f_0'\left( 1 + \frac{3\Omega'^2}{2{\omega'_0}^2}\right) + \mathcal{O} \left(\Omega'^4, \Omega'^2A'^2, A'^4 \right),
\end{equation}
and the center position of the oscillation, $\xi'_{\rm center}$, is obtained as
\begin{equation}
\xi'_{\rm center} = \frac{\Omega'^2}{{\omega_0}'^2} + \mathcal{O} \left(\Omega'^4 ,\Omega'^2A'^2, A'^4 \right).
\end{equation}
where $\omega'_0 = 2 \pi f'_0 = \sqrt{2k/m}$ is the angular frequency without rotation, $\Omega' = \sqrt{J/(2m{\ell_0}^2)}$ corresponds to the parameter for the frequency of the rotation, and $A'$ denotes the amplitude of the oscillation. The center-position shift is calculated as zero when the terms with the same order as $\Omega'^2$ and $A'^2$ are taken into consideration.
From the results, the increase in frequency when a droplet is rotating can be considered to be due to the surface tension, which works like a spring. 
However, a spring-bead model with a linear spring cannot reproduce the center-position shift and the frequency dependence on the amplitude.

By taking into account the nonlinearity of the spring, {\it i.e.}, $U(\ell)$ is set as $U(\ell) = k(\ell - 2\ell_0)^2/2 + k_2(\ell - 2\ell_0)^3/3 + k_3(\ell - 2\ell_0)^4/4 + k_4(\ell - 2\ell_0)^5/5$, the frequency, $f'$, and center-position shift, $\xi'_{\rm center}$, are calculated as
\begin{align}
f' =& f'_0 \left(1 + \left(\frac{3}{2} + \frac{K_2}{{\omega_0'}^2}\right)\frac{\Omega'^2}{{\omega_0'}^2} \right. \nonumber \\ &\left.+ \left(- \frac{{K_2}^2}{2 {\omega'_0}^4} + \frac{3K_3}{8{\omega_0'}^2}\right)A'^2 + \mathcal{O} \left(\Omega'^4, \Omega'^2A'^2, A'^4 \right)\right),
\end{align}
\begin{equation}
\xi'_{\rm center} = \frac{\Omega'^2}{{\omega_0}'^2} - \frac{K_2}{2{\omega_0}'^2} A'^2 + \mathcal{O} \left(\Omega'^4 ,\Omega'^2A'^2, A'^4 \right).
\end{equation}
Here, $K_2 = 4 k_2 \ell_0/m$ and $K_3 = 8 k_3 {\ell_0}^2 / m$. 
By setting $K_2 > 0$ and ${K_2}^2 > 3 {\omega_0'}^2 K_3 / 4$, the spring-bead model can qualitatively reproduce the frequency shift and center-position shift with respect to the amplitude and rotational angular velocity. It should be noted that these conditions cannot be realized when only the surface energy of the droplet is considered as the potential in eq.~(\ref{surface}). The detailed calculations are shown in Appendix~\ref{ap-comparison}.

\section{Numerical Calculation}

\begin{figure}
\begin{center}
\includegraphics{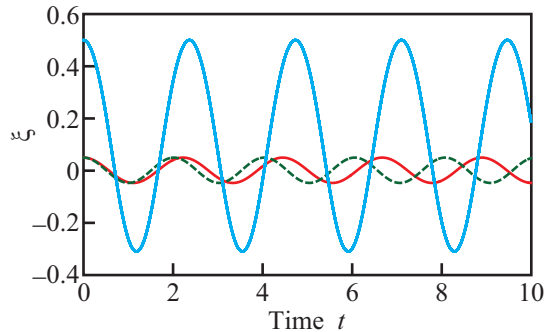}
\end{center}
\caption{(Color online) Time series of $\xi$ calculated based on eq.~(\ref{xi}). The solid red (dark gray) and broke green (dark gray) curves correspond to the conditions without and with rotation, respectively. The solid cyan (light gray) curve shows the results for larger initial amplitude. The center position of the oscillation for the larger amplitude shifted to a positive value of $\xi$. The parameters and initial conditions, $\xi_{\rm ini}$, are set as (solid red curve) $\Omega = 0, \xi_{\rm ini} = 0.05$, (broken green curve) $\Omega = 1, \xi_{\rm ini} = 0.05$, and (solid cyan curve) $\Omega = 0, \xi_{\rm ini} = 0.5$. The other parameters are the same for all the plots: $\gamma = \rho = R = 1$.}
\label{fig2}
\end{figure}

In order to confirm the analytical results, numerical calculations were carried out based on eq.~(\ref{xi}). The Euler method was used with a time step of $10^{-5}$. First, the effects of rotation and amplitude change were evaluated. The time series of $\xi$ for different rotational angular velocities and amplitudes are shown in Fig.~\ref{fig2}. From the numerical results, it was confirmed that the frequency increases when the droplet is rotating and that the frequency decreases as the amplitude increases.

\begin{figure}
\begin{center}
\includegraphics{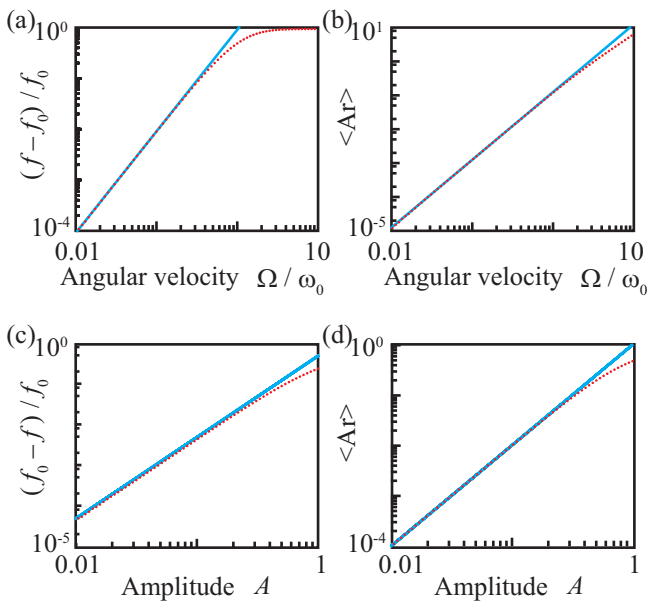}
\end{center}
\caption{(Color online) Frequency shift and averaged aspect ratio with respect to rotational angular velocity and oscillation amplitude. (a) Frequency shift with respect to angular velocity. (b) Averaged aspect ratio with respect to angular velocity. (c) Frequency shift with respect to amplitude. (d) Averaged aspect ratio with respect to the amplitude. All the plots are in the logarithmic scale. The results of the numerical calculation are shown by the red (dark gray) points, whereas the cyan (light gray) lines correspond to the analytical results from eqs.~(\ref{freq}) and (\ref{ar}).}
\label{fig3}
\end{figure}

In order to compare the numerical results with the analytical predictions, the frequency shift, $(f-f_0)/f_0$, and averaged aspect ratio, $\left< {\rm Ar} \right>$, with respect to the rotational angular velocity and oscillation amplitude were calculated. The method used for the numerical calculation was the same as above, and the results are shown in logarithmic plots in Fig.~\ref{fig3}. The numerical results correspond well with the analytical ones.

\section{Comparison with Experiments}

In order to confirm the validity of the proposed model, the analytical results were compared with experimental ones. 
A brief description of the experimental setting is provided below. A more detailed explanation can be found in another paper~\cite{Tanaka}. 
\begin{figure}
\begin{center}
\includegraphics{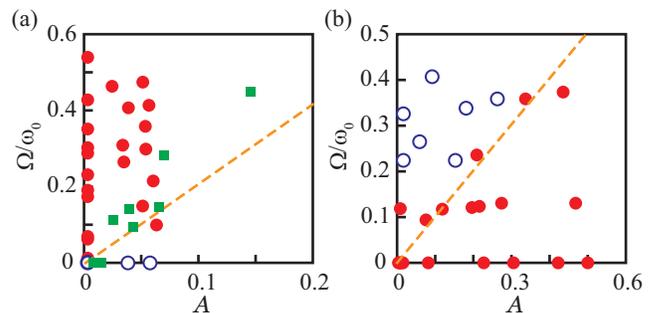}
\end{center}
\caption{(Color online) Comparison of analytical and experimental results. (a) Frequency shift of oscillation in ellipsoidal deformation with respect to nondimensionalized rotational angular velocity, $\Omega/\omega_0$, and nondimensionalized amplitude, $A$. The blue (open) and red (closed) circles represent the positive and negative frequency shifts, respectively, observed in the experiments, and the green (light gray) squares represent zero frequency shifts. The condition for zero frequency shift obtained analytically is shown by the broken line (eq.~(\ref{eq-fr})). (b) Averaged aspect ratio of droplet with respect to nondimensionalized rotational angular velocity, $\Omega/\omega_0$, and nondimensionalized amplitude, $A$. The blue (open) and red (closed) circles represent the prolate- and oblate-like deformations, respectively, in the averaged shape observed in the experiments. The condition for the averaged aspect ratio to be one obtained analytically, is shown by the broken line (eq.~(\ref{eq-as})).}
\label{fig4}
\end{figure}
\begin{figure}
\begin{center}
\includegraphics{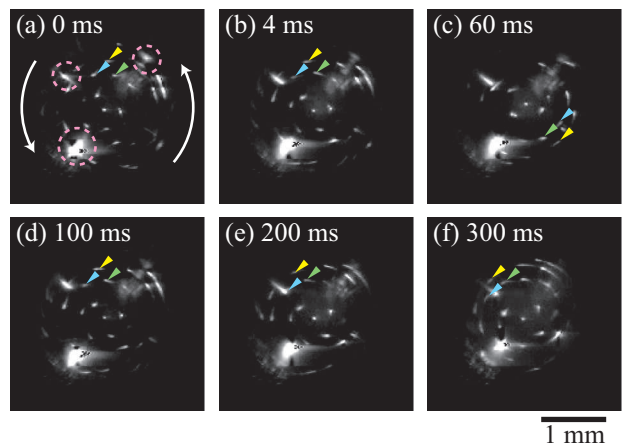}
\end{center}
\caption{(Color online) Results of experiments to check whether the droplet rotates like an elastic rigid body. Snapshots taken from the direction of the rotational axis are shown. The time shown in the figure represents the elapsed time, and each arrow indicates the position of a tracer particle. The droplet was rotating counterclockwise, and the rotational frequency was approximately 10.5~Hz (angular velocity is 66~rad/s), which corresponds to a period of 95~ms. In this case, the droplet exhibited oscillations in the ellipsoidal deformation at $\sim$ 70.9~Hz. The three regions surrounded by the broken circles in (a) were bright due to the direct reflection of the light illuminated for visualization of small particles. The white arrows in (a) indicate the direction of rotation.}
\label{fig5}
\end{figure}

In order to levitate the droplet, a pair of electrodes were set up such that one electrode was located just above the other. By applying a positive voltage at the lower electrode, the droplet injected from a syringe connected to the lower electrode was electrified. The volume of the droplet could be adjusted by
controlling the volume of the injected liquid. After injection, negative voltage was applied at the upper electrode, and the droplet was levitated. Then, an additional sinusoidal voltage was applied at the lower electrode in order to induce an oscillation in the ellipsoidal deformation of the droplet. By scanning the frequency of the applied sinusoidal voltage, the resonance frequency of the droplet was obtained. In order to realize the rotation of the levitated droplet, standing acoustic waves were generated in a container using a pair of orthogonally positioned acoustic drivers so that a torque was exerted on the droplet~\cite{acousticrotation}. The rotational angular velocity was controlled by adjusting the sound pressure.

The droplet was illuminated by a He-Ne laser, and its vertical position was determined using a position detector. The position of the levitated droplet was controlled by changing upper electrode voltage, which was determined by the position feedback of a proportional-integral-derivative (PID) algorithm. The droplet radius was measured using a line sensor, and the time change in the shape of the droplet was recorded using two high-speed cameras. 
For the observation of the flow field inside the droplet, tracer particles made of nylon (9~-~13~$\mu {\rm m}$ in diameter) were dispersed into the droplet and observed from above.
The test liquid used for generating the droplet was propylene carbonate, whose density, surface tension, and viscosity are 1206~${\rm kg}/{\rm m}^3$, 43~${\rm mN/m}$, and 2.7~${\rm mPa} \cdot {\rm s}$, respectively, at room temperature.

From eq.~(\ref{freq}), the relationship between the nondimensionalized rotational angular velocity, $\Omega/\omega_0$, and the nondimensionalized amplitude, $A$, for the condition in which the frequency does not shift significantly compared with that in the case of the sufficiently small deformation without rotation can be obtained as
\begin{equation}
\frac{\Omega}{\omega_0} = \sqrt{\frac{1037}{2128}} A. \label{eq-fr}
\end{equation}
Moreover, the relationship between $\Omega/\omega_0$ and $A$ for the condition in which the averaged aspect ratio is one can be obtained as
\begin{equation}
\frac{\Omega}{\omega_0} = \sqrt{\frac{57}{56}} A. \label{eq-as}
\end{equation}
The experimental observations for the  different rotational angular velocities and oscillation amplitudes are compared with the analytical results in Fig.~\ref{fig4}.

In order to confirm the assumptions used in constructing the simple model, it was experimentally verified whether the droplet rotates like an elastic rigid body and whether the angular momentum is conserved.
The velocity profile of the fluid inside the droplet was obtained from the motion of the tracer particles observed from above, {\it i.e.}, in the direction of the rotational axis. From the recorded video, the position of the rotational center and the time series of the positions of the tracer particles were determined. The configuration of the particles did not change much as shown in Fig.~\ref{fig5}, in which three particles were traced.
\begin{figure}
\begin{center}
\includegraphics{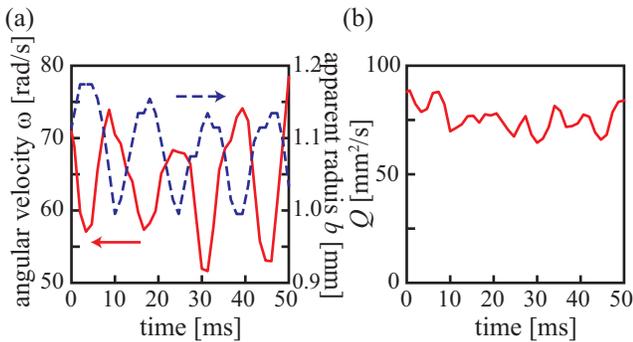}
\end{center}
\caption{(Color online) (a) Time evolution of angular velocity of the droplet, $\omega$ (solid red curve, left vertical axis), and apparent radius, $b$ (broken blue curve, right vertical axis). (b) Time evolution of angular momentum per mass, $Q$, defined as $Q = b^2 \omega$.}
\label{fig6}
\end{figure}

Furthermore, we also measured the angular velocity of the particle, $\omega$, and the apparent radius, {\it i.e.}, the radius of the droplet in the plane perpendicular to the $z$-axis, $b$, with respect to time were obtained from the video as shown in Fig.~\ref{fig6}(a). It was found that the angular velocity oscillates in synchronization with the change in $b$. In addition, the angular momentum per mass, $Q$, was defined as $Q = b^2 \omega$ and plotted with respect to time in Fig.~\ref{fig6}(b). The change in $Q$ was approximately 20\% of the averaged value. From these results, it can be concluded that the assumptions (ii) and (iii), {\it i.e.}, fixed configuration inside the droplet and angular momentum conservation due to the small energy dissipation, are adequate.

\section{Summary}

A droplet of the millimeter-to-centimeter scale can be levitated by applying an electric charge, and such levitated droplets have been used to measure the surface tension of liquids by determining the characteristic frequency of oscillatory deformation. In the present study, a simple mechanical model was proposed by considering a single oscillation mode of the ellipsoidal deformation of a levitated rotating droplets. By measuring the dependence of the frequency on the amplitude and rotational speed, it was expected that the accuracy of the surface tension measurement could be improved. By the proposed model, the characteristic frequency of the oscillation for a rotating droplet can be calculated. Moreover, the dependence of the frequency on the rotating speed and amplitude observed in the experiment was reproduced by the model.

\begin{acknowledgments}
 This work was supported in part by Grants-in-aid for Scientific Research (B) (No.15H03925) and for Scientific Research on Innovative Areas ``Fluctuation \& Structure'' (No.25103008) and the Core-to-Core Program ``Nonequilibrium dynamics of soft matter and information'' to H.K. from the Japan Society for the Promotion of Science (JSPS).
\end{acknowledgments}

\appendix

\section{Confirmation of flow field as solution of Navier-Stokes equation}
\label{ap-NS}

In the main text, the velocity field inside the ellipsoidal droplet was assumed as
\begin{equation}
{\bf v} = \frac{\dot{b} r}{b} {\bf e}_r + r \dot{\theta} {\bf e}_\theta + \frac{\dot{a} z}{a} {\bf e}_z = - \frac{\dot{a}r}{2a} {\bf e}_r + \frac{\Omega a r}{R}  {\bf e}_\theta + \frac{\dot{a} z}{a}  {\bf e}_z, \label{flowfield}
\end{equation}
in the cylindrical coordinates. 

In this appendix, it is shown that the flow field in eq.~(\ref{flowfield}) satisfies the Navier-Stokes equation
\begin{equation}
\rho \left( \frac{\partial}{\partial t} + {\bf v} \cdot \nabla \right){\bf v} = \eta \nabla^2 {\bf v} - \nabla P, \label{NS}
\end{equation}
with incompressibility
\begin{equation}
\nabla \cdot {\bf v} = 0, \label{incompressible}
\end{equation}
where $\rho$ is the density of the fluid, $\eta$ is the viscosity of the fluid, $P$ is the pressure,  and ${\bf e}_r$, ${\bf e}_\theta$, and ${\bf e}_z$ are the unit vectors in the $r$, $\theta$, and $z$ directions, respectively. It is assumed that no external force is exerted on the droplet.

In the cylindrical coordinates, the Navier-Stokes equation can be written as
\begin{widetext}
\begin{equation}
\rho \left( {\frac{{\partial {v_r}}}{{\partial t}} + {v_r}\frac{{\partial {v_r}}}{{\partial r}} + \frac{{{v_\theta }}}{r}\frac{{\partial {v_r}}}{{\partial \theta }} + {v_z}\frac{{\partial {v_r}}}{{\partial z}} - \frac{{v_\theta ^2}}{r}} \right)  =  - \frac{{\partial P}}{{\partial r}} + \eta \left( {\frac{{{\partial ^2}{v_r}}}{{\partial {r^2}}} + \frac{1}{r}\frac{{\partial {v_r}}}{{\partial r}} + \frac{1}{{{r^2}}}\frac{{{\partial ^2}{v_r}}}{{\partial {\theta ^2}}} + \frac{{{\partial ^2}{v_r}}}{{\partial {z^2}}} - \frac{2}{{{r^2}}}\frac{{\partial {v_\theta }}}{{\partial \theta }} - \frac{{{v_r}}}{{{r^2}}}} \right),
\end{equation}
\begin{equation}
 \rho \left( {\frac{{\partial {v_\theta }}}{{\partial t}} + {v_r}\frac{{\partial {v_\theta }}}{{\partial r}} + \frac{{{v_\theta }}}{r}\frac{{\partial {v_\theta }}}{{\partial \theta }} + {v_z}\frac{{\partial {v_\theta }}}{{\partial z}} + \frac{{{v_r}{v_\theta }}}{r}} \right)= - \frac{1}{r}\frac{{\partial P}}{{\partial \theta }} + \eta \left( {\frac{{{\partial ^2}{v_\theta }}}{{\partial {r^2}}} + \frac{1}{r}\frac{{\partial {v_\theta }}}{{\partial r}} + \frac{1}{{{r^2}}}\frac{{{\partial ^2}{v_\theta }}}{{\partial {\theta ^2}}} + \frac{{{\partial ^2}{v_\theta }}}{{\partial {z^2}}} + \frac{2}{{{r^2}}}\frac{{\partial {v_r}}}{{\partial \theta }} - \frac{{{v_\theta }}}{{{r^2}}}} \right),
\end{equation}
\begin{equation}
\rho \left( {\frac{{\partial {v_z}}}{{\partial t}} + {v_r}\frac{{\partial {v_z}}}{{\partial r}} + \frac{{{v_\theta }}}{r}\frac{{\partial {v_z}}}{{\partial \theta }} + {v_z}\frac{{\partial {v_z}}}{{\partial z}}} \right) = - \frac{{\partial P}}{{\partial z}} + \eta \left( {\frac{{{\partial ^2}{v_z}}}{{\partial {r^2}}} + \frac{1}{r}\frac{{\partial {v_z}}}{{\partial r}} + \frac{1}{{{r^2}}}\frac{{{\partial ^2}{v_z}}}{{\partial {\theta ^2}}} + \frac{{{\partial ^2}{v_z}}}{{\partial {z^2}}}} \right),
\end{equation}
\end{widetext}
for each component, where the flow profile is represented in the cylindrical coordinates as ${\bf v} = v_r {\bf e}_r + v_\theta {\bf e}_\theta+ v_z {\bf e}_z$~\cite{Landau}.
By setting the pressure, $P$, as 
\begin{equation}
P = P_0 + \rho \left( \frac{1}{4}\frac{\ddot{a}}{a} r^2- \frac{3}{8} \left(\frac{\dot{a}}{a}\right)^2 {r^2}- \frac{a^2 \Omega^2}{2R^2}r^2 - \frac{1}{2}\frac{\ddot{a}}{a} z^2 \right), \label{pressure}
\end{equation}
the flow profile satisfies the Navier-Stokes equation in the cylindrical coordinates. Here, the volume conservation and the equality, $\dot{\theta}/a = \Omega / R$, which is derived in the main text, were used.

The incompressibility condition (eq.~\ref{incompressible}) also holds: 
\begin{equation}
\nabla \cdot {\bf v} = \frac{{\partial {v_r}}}{{\partial r}} + \frac{{{v_r}}}{r} + \frac{1}{r}\frac{{\partial {v_\theta }}}{{\partial \theta }} + \frac{{\partial {v_z}}}{{\partial z}} = 0.
\end{equation}

\section{Derivation of shifts in frequency and center position with respect to amplitude}
\label{ap-derivation}

In this section, the shifts in frequency and center position are derived by comparing with a harmonic oscillator. The system includes the nonlinear terms as 
\begin{equation}
\ddot{x} + {\omega_0}^2 x + \alpha_1 \dot{x}^2 + \alpha_2 x \dot{x}^2 + \alpha_3 x^2 + \alpha_4 x^3 = 0, \label{nonlin}
\end{equation}
where $\ddot{x}$ and ${\omega_0}^2 x$ are the main terms and the other terms are sufficiently small.
By letting $x = A \cos p t + B$, where $A$ and $B$ are of the order of $\epsilon$, and substituting in eq.~(\ref{nonlin}), 
\begin{widetext}
\begin{align}
\ddot{x} + p^2 (x - B) =& (p^2 - \omega^2) (A \cos pt + B) - \alpha_1 A^2 p^2 \sin^2 pt - \alpha_2 A^2 p^2 \sin^2 pt (A \cos pt + B) \nonumber \\ &- \alpha_3 (A \cos pt + B)^2 - \alpha_4 (A \cos pt + B)^3 - B p^2 \nonumber \\ 
=& - \omega^2 B - \frac{\alpha_1}{2} A^2 p^2 - \frac{\alpha_2}{2} A^2 B p^2  - \alpha_3 B^2 - \frac{\alpha_3}{2} A^2 - \alpha_4 B^3 - \frac{3 \alpha_4}{2} A^2 B \nonumber \\
& + \left[(p^2 - \omega^2)A - \frac{\alpha_2}{4}A^3 p^2 - 2 \alpha_3 A B - 3 \alpha_4 A B^2 - \frac{3 \alpha_4}{4} A^3 \right] \cos pt \nonumber \\
& + \left[ \frac{\alpha_1}{2} A^2 p^2 + \frac{\alpha_2}{2} A^2 B p^2 - \frac{\alpha_3}{2} A^2 - \frac{3\alpha_4}{2} A^2 B \right] \cos 2pt + \left[ \frac{\alpha_2}{4} A^3 p^2 - \frac{\alpha_4}{4} A^3 \right] \cos 3pt.
\end{align}
\end{widetext}
The constant term and the secular term have to be necessarily zero for the system to exhibit oscillation.
Considering the balance of the infinitesimally small terms, $B$ is of the same order as $A^2$. Therefore,
\begin{equation}
B = -\frac{\alpha_1}{2 \omega^2}A^2 p^2 - \frac{\alpha_3}{2 \omega^2}A^2,
\end{equation}
which gives
\begin{equation}
p^2 - \omega^2 = \left( -\alpha_1 \alpha_3 + \frac{1}{4} \alpha_2 \omega^2 - \frac{1}{\omega^2} {\alpha_3}^2 + \frac{3}{4} \alpha_4\right) A^2.
\end{equation}
Thus, the frequency, $p$, is calculated as
\begin{equation}
\frac{p}{\omega} =1 + \left( -\frac{1}{2} \alpha_1 \frac{\alpha_3}{\omega^2} + \frac{1}{8} \alpha_2 - \frac{1}{2} \left( \frac{\alpha_3}{\omega^2}\right)^2 + \frac{3}{8} \frac{\alpha_4}{\omega^2} \right) A^2, \label{eq-freq}
\end{equation}
and the shift in the center position, $B$, is calculated as
\begin{equation}
B = -\left( \frac{\alpha_1}{2} + \frac{\alpha_3}{2\omega^2}\right) A^2. \label{eq-shift}
\end{equation}

\section{Comparison with simple spring-bead model}
\label{ap-comparison}

\begin{figure}
\begin{center}
\includegraphics{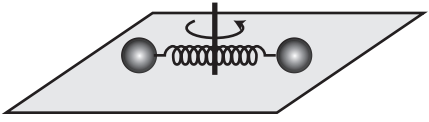}
\end{center}
\caption{Schematic illustration of the simple spring-bead model.}
\label{fig7}
\end{figure}
A system with two equivalent small beads connected by a spring in a two-dimensional plane is considered as shown in Fig.~\ref{fig7}. A force is exerted on the two beads as a result of the potential corresponding to the extension of spring, $U(r)$. The Lagrangian of the system can be written as
\begin{equation}
\mathcal{L} = \frac{m}{2} \left ( {\dot{r}_1}^2 + {\dot{r}_2}^2 + {r_1}^2 {\dot{\theta}_1}^2 + {r_2}^2 {\dot{\theta}_2}^2 \right ) - U(\left|{\bf r}_2 - {\bf r}_1\right|),
\end{equation}
where $m$ is the mass of the bead,  ${\bf r}_i$ is the positional vector of the $i$-th bead ($i = 1, 2$), and $r_i$ and $\theta_i$ are the components of ${\bf r}_i$ in the two-dimensional polar coordinates. The potential energy of the spring can be written as
\begin{align}
U(r) =& \frac{k}{2} \left(r - 2\ell_0 \right)^2 + \frac{k_2}{3} \left(r - 2\ell_0 \right)^3  \nonumber \\ &+ \frac{k_3}{4} \left(r - 2 \ell_0 \right)^4 + \frac{k_4}{5} \left(r - 2 \ell_0 \right)^5.
\end{align}
The Euler-Lagrange equation for $\theta_i$ represents the angular momentum conservation, {\it i.e.}, $m {r_i}^2 \dot{\theta_i} = J_i$. By assuming that the center of mass of the system is always at the origin and the oscillation always occurs in a symmetric manner, {\it i.e.}, $r_1 = r_2 = r$, and $\theta_1 = \theta_2 + \pi = \theta$, it can be obtained that $2mr^2 \dot{\theta} = J$, where $J$ is the initial angular momentum. 

Then, the governing equation is obtained as
\begin{align}
\ddot{r} =& \frac{J^2}{4m^2} \frac{1}{r^3} - \frac{2k}{m} (r -\ell_0) - \frac{4{k_2}}{m} (r - \ell_0 )^2 \nonumber \\ & -\frac{8{k_3}}{m} (r - {\ell_0})^3 -\frac{16{k_4}}{m} (r - {\ell_0})^4,
\end{align}
where $r_1$ and $r_2$ are substituted by $r$.

Without rotation, the equilibrium point is $r = \ell_0$, and therefore the position is normalized by $\ell_0$, {\it i.e.}, the nondimensionalized variable $\xi'$ is set as $\xi' = (r -\ell_0) / \ell_0$. Then
\begin{equation}
\ddot{\xi} = {\Omega'}^2\frac{1}{ (1 + \xi')^3}- {{\omega'}_0}^2 \xi' - K_2 \xi'^2 - K_3 \xi'^3 - K_4 \xi'^4.
\end{equation}
Here, $\omega'_0$ and $\Omega'$ are set as
\begin{equation}
\omega'_0 = \sqrt{\frac{2k}{m}},
\end{equation}
\begin{equation}
\Omega' = \sqrt{\frac{J}{2 m {\ell_0}^2}},
\end{equation}
and the normalized values $K_2$, $K_3$, and $K_4$ are set as
\begin{equation}
K_2 = \frac{4k_2 \ell_0}{m},
\end{equation}
\begin{equation}
K_3 = \frac{8k_3 {\ell_0}^2}{m},
\end{equation}
\begin{equation}
K_4 = \frac{16k_4 {\ell_0}^3}{m},
\end{equation}

The equilibrium point is
\begin{equation}
\xi'_0 = \frac{{\Omega'}^2}{{{{\omega'}_0}}^2} + \mathcal{O} \left({\Omega'}^4 \right).
\end{equation}
By expanding the governing equation with respect to $\xi'_0$, the 
frequency is obtained as
\begin{align}
f' =& f'_0 \left( 1 + \left(\frac{3}{2} + \frac{K_2}{{\omega'_0}^2}\right) \frac{{\Omega'}^2}{{\omega'_0}^2} + \left(-\frac{{K_2}^2}{2 {\omega'_0}^4} + \frac{3K_3}{8{\omega'_0}^2} \right) {A'}^2\right) \nonumber \\ &+ \mathcal{O} \left( {\Omega'}^4, {\Omega'}^2{A'}^2, A'^4 \right), 
\end{align}
and the center-position shift, which corresponds to the average value of $b$, is determined as 
\begin{equation}
\xi'_{\rm center} = \frac{{\Omega'}^2}{{\omega'_0}^2} - \frac{K_2}{2{\omega'_0}^2} {A'}^2 + \mathcal{O} \left( {\Omega'}^4, {\Omega'}^2{A'}^2, A'^4 \right).
\end{equation}
In this calculation, the results from Appendix~\ref{ap-derivation} were used.

\end{document}